\documentclass[10pt,conference]{IEEEtran}

\usepackage{cite}
\usepackage{graphicx}
\usepackage{xcolor}
\usepackage{tabularx}
\usepackage{url}
\usepackage{flushend}
\usepackage{booktabs}
\usepackage[numbers]{natbib}

\usepackage{multirow}
\usepackage{adjustbox}

\usepackage{listings}
\definecolor{dkgreen}{rgb}{0,0.6,0}
\definecolor{gray}{rgb}{0.5,0.5,0.5} 
\definecolor{mauve}{rgb}{0.58,0,0.82} 

\usepackage{tcolorbox}

\begin{document}

\title{Atoms of Confusion in Java}

\author{\IEEEauthorblockN{Chris Langhout}
\IEEEauthorblockA{\textit{Software Engineering Research Group} \\
\textit{Delft University of Technology}\\
Delft, The Netherlands \\
chris.langhout@gmail.com}
\and
\IEEEauthorblockN{Maurício Aniche}
\IEEEauthorblockA{\textit{Software Engineering Research Group} \\
\textit{Delft University of Technology}\\
Delft, The Netherlands \\
M.FinavaroAniche@tudelft.nl}
}

\maketitle

\begin{abstract}
Although writing code seems trivial at times, problems arise when humans misinterpret what the code actually does.
One of the potential causes are ``atoms of confusion'', the smallest possible patterns of misinterpretable source code.
Previous research has investigated the impact of atoms of confusion in C code. Results show that developers make significantly more mistakes in code where atoms are present.
In this paper, we replicate the work of Gopstein et al. to the Java language. After deriving a set of atoms of confusion for Java, we perform a two-phase experiment with 132 computer science students (i.e., novice developers). 
Our results show that participants are 2.7 up to 56 times more likely to make mistakes in code snippets affected by 7 out of the 14 studied atoms of confusion, and when faced with both versions of the code snippets, participants perceived the version affected by the atom of confusion to be more confusing and/or less readable in 10 out of the 14 studied atoms of confusion.
\end{abstract}


\newcommand{\RQone}{Which atoms of confusion hinder the comprehensibility of Java programs, and to what extent?}
\newcommand{\RQtwo}{How do students perceive confusion in Java programs that include atoms of confusion, as opposed to the translated, confusion-free, Java programs?}

\section{Introduction}
\label{cha:intro}

Writing source code in such a way that developers effectively understand it is fundamental for the sustainable development, maintenance, and evolution of software systems~\cite{gopstein2017understanding, Schroter2017, Siegmund2016}.
In contrast to natural languages, programming languages have an unambiguous meaning for a valid syntactical piece of code~\cite{Avidan2017}.
However, developers may not always draw the correct conclusions on the behaviour of a piece of code; they can mistake the meaning of code and misjudge the program's behaviour, leading to errors~\cite{gopstein2017understanding}.

Different programming languages give software developers many ways of implementing a solution to a given problem. 
For example, the simple task of converting a boolean \texttt{true} or \texttt{false} value into a numeric integer value can be coded in a vast amount of ways.
An example can be found in the difference in the answers given to the question \textit{How to convert boolean to int in Java?} in StackOverflow\footnote{\url{https://stackoverflow.com/questions/3793650/convert-boolean-to-int-in-java}. Accessed in July 25, 2019}.
We display eight different answers in Figure~\ref{fig:SOExamples}, ordered by their amount of votes. 
The solutions show a significant amount of variation in logic, readability, and understandability.
Using the \texttt{ternary if} operator (option 1) is considered the `most readable' by the commenters, and received the most votes, making it the most accepted solution according to the rules of the StackOverflow community.
Interestingly, this is in contrast with the findings of \citet{gopstein2017understanding}, which show that the use of the \textit{conditional operator} atom is found to be significantly confusing.
Furthermore, the author of the 8th answer starts their answer with ``\textit{If you want to obfuscate, use this:}'' showing that the intention of this answer is not readability, but instead showing a less known alternative to solve the question.

Discussions about code misconceptions happen regularly.
On June 14th, 2019, Jonathan Wakely started a discussion on the bug tracker of GCC about introducing warnings when people use the boolean operator \(\wedge{}\) with integer literals.\footnote{\url{https://gcc.gnu.org/bugzilla/show_bug.cgi?id=90885}}
The \(\wedge{}\) operator represents a bitwise XOR operation in most programming languages.
Instead, developers confuse the symbol with the mathematical representation of power.
The author states: ``\textit{There's nothing wrong about implicit fallthrough, misleading indentation, ambiguous else, or missing parentheses in nested logic expressions either. But people get it wrong all the time.
I can't see a good reason to write 2\(\wedge{}\)16 when you mean 18, or 10\(\wedge{}\)9 when you mean 3, so it's probably a bug. And there's an easy workaround to avoid the warning: just write the exact constant as a literal, not an XOR expression.}''\footnote{\label{fn:SOURL}\url{https://gcc.gnu.org/bugzilla/show_bug.cgi?id=90885#c6}}
Other developers jumped in the discussion to show examples of occurrences of this particular pattern on GitHub and other source code hosting sites.
Several of the responses provide suggestions to specific cases when a warning should, or should not be raised, depending on the use of literals or not.
These examples show that readable code is a relevant topic, and that the details up to the smallest code snippets can make a difference in understandability.

\citet{gopstein2017understanding} observed a trend in notable software bug examples, where the failure is caused by \textit{``a single, well-contained, programming error at the syntactic or semantic level, rather than the algorithmic or system-levels of the project''}. In this work, authors explored the idea of \textbf{atoms of confusion}, or ``atoms'' for short, which they define as minimal portions of code that cause a person to different conclusions on the output.
\citet{castor2018identifying} expanded this definition by formalizing atoms as precisely identifiable, likely to cause confusion, replaceable by a functionally equivalent code pattern that is less likely to cause confusion, and indivisible.
Atoms of confusion do not include non-deterministic, undefined/non-portable, computational, and API-related code, since the target of the atoms is the mistakes caused by misunderstanding.

In that work, \citet{gopstein2017understanding} also show a significant increase in misunderstanding caused by the C atoms of confusion patterns, opposed to code without the atoms of confusion in an experiment with 73 participating students. 
To show the impact of these confusion patterns, authors also performed a second experiment, with 43 participants and more extensive confusing programs.
The results of this second experiment show statistically significantly higher error rates in the evaluation of obfuscated variants of programs.

The goal of this paper is to generalize the knowledge on atoms of confusion to the Java programming language and gain insights in what makes these atoms confusing.
To that aim, we propose a two-step experiment with 132 computer science students (i.e., novice developers).
First, we evaluate the impact of the atoms of confusion through an experiment where we randomly show code snippets, that may or may not contain an atom of confusion, to participants and ask them to evaluate the output of the program (in a design similar to the original research).
Second, we measure the participants' preferences by presenting both the confusing and the non-confusion versions of the same code snippet to participants, and asking them to indicate which one they perceive as more confusing.

Our results show that (i) participants are 2.7 up to 56 times more likely to make mistakes in code snippets affected by 7 out of the 14 studied atoms of confusion, and (ii) when faced with both versions of the code snippets, participants perceived the version affected by the atom of confusion to be more confusing in 10 out of the 14 studied atoms of confusion.

The contributions of this work are:

\begin{enumerate}
    \item A set of atoms of confusion for the Java language, based on the atoms of confusion proposed by \citet{gopstein2017understanding}.
    \item An empirical evaluation of the impact and the perception of the Java atoms of confusion among 132 novice developers (i.e., computer science students).
\end{enumerate}

\begin{figure}
    \centering
\begin{lstlisting}
1. i = b ? 1 : 0;\end{lstlisting}\vspace{2\baselineskip{}}
\begin{lstlisting}
2. i = (Boolean b).compareTo(false);\end{lstlisting}\vspace{2\baselineskip{}}
\begin{lstlisting}
3. i = Boolean.compare(b, false);\end{lstlisting}\vspace{2\baselineskip{}}
\begin{lstlisting}
4. i = -("false".indexOf(""+b));\end{lstlisting}\vspace{2\baselineskip{}}
\begin{lstlisting}
5. i = 5 - b.toString().length;\end{lstlisting}\vspace{2\baselineskip{}}
\begin{lstlisting}
6. import org.apache.commons.lang3.
                    BooleanUtils;
    i = BooleanUtils.toInteger(b);\end{lstlisting}\vspace{2\baselineskip{}}
\begin{lstlisting}
7.  if(b){
        i = 1;
    } else{
        i = 0;
    }\end{lstlisting}\vspace{2\baselineskip{}}
\begin{lstlisting}
8. i = 1 & Boolean.hashCode( b ) >> 1;\end{lstlisting}
    \vspace{.2cm}
    \caption[Examples from StackOverflow answers on how to convert boolean \texttt{b} to int in Java]{Examples from StackOverflow answers on how to convert boolean \texttt{b} to an int in Java~\textsuperscript{\ref{fn:SOURL}}}\label{fig:SOExamples}
\end{figure}
\section{Background and Related Work}

Program comprehension is a widely explored domain in computer science.
For this study, we are mainly interested in the skill level in program comprehension of novice programmers.

In 1985, \citet{bonar1985preprogramming} stated that \textit{`many programming bugs can be explained by novices inappropriately using their knowledge of step-by-step procedural specifications in natural language'}.
We can take away that certain bugs can be caused by lack of expertise. Finding out what programming code does, or what code is needed, our intuition sometimes tries to connect our present knowledge from other fields to find a solution.
In that particular research, knowledge of natural language tricked participants into making mistakes when writing code.

\citet{Ajami2017} use an experimental platform fashioned as an online game-like environment to measure how quickly and accurately 220 professional programmers interpret code snippets with similar functionality but different structures.
The findings include that there is no relation between errors made and time taken to understand the snippets, but snippets that take longer to understand are considered harder~\cite{Ajami2017}.
When a snippet contains a \texttt{for} loop, the code is considered much harder to understand compared to snippets containing \texttt{if}s.
Snippets with predicates become harder to understand when negations are present, and \texttt{for} loops counting down are harder to understand than loops that count up.
This shows that the slight differences in the way of expressing predicates can be measured when compared against the use of known idioms.
The syntactic structures of code are shown to not necessarily take up the biggest part in the measurement of complexity of code.
Authors also found that the metrics of time to understanding and the amount of errors made are not necessarily related.
This means that the amount of errors made is not related to how long a participant takes to solve a problem.


By understanding what makes code less readable, we can draw conclusions about what code constructs are understandable and which to avoid.
This is shown by the research of \citet{gopstein2017understanding}.
The main source of the code patterns used for the atoms of confusion comes from a contest on writing obfuscated code called IOCCC (the International Obfuscated C Code Contest).
Authors describe 19 atoms of confusion, on which this research is based on.
To show the real-world relevance of these selected atoms of confusion, follow up research from \citet{gopstein2018prevalence} shows that the 15 atoms that were proven to be confusing, occur in practice once per 23 lines.
Their research is based on the analysis of 14 of the most popular and influential \texttt{C} and \texttt{C++} software projects.
\citet{medeiros2018investigation} also researched the rate of occurrences of most of the atoms and show that all but one occur in the analyzed projects.
They based their numbers on a set of 50 open-source \texttt{C} projects using a mixed method approach including repository mining and developer surveys.
Four of the 12 atoms researched by \citet{medeiros2018investigation} are shown to be commonly used.

In a more recent work, Gopstein et al.~\cite{gopstein-fse} performed a think-aloud study in which researchers observe 14 developers (students and professional developers) as they hand-evaluated confusing code (i.e., code containing Gopstein's previously defined atoms of confusion). The results show that atoms may confuse developers for many different reasons, i.e., developers are not familiarized with that code construct, 
or are familiarized with it but attributes incorrect semantics to it, or even lack of attention when evaluating the entire snippet.

A similar approach to researching patterns of code that cause misinterpretation can be found in the work of \citet{Dolado2003}. 
This work provides insights in misinterpretations caused by code that has side-effects.
The researched code fragments are comparable to code examples used in this study. The atoms \textit{pre increment decrement}, \textit{post increment decrement} and \textit{logic as control flow} have most similarities, as they also make use of expressions with side-effects.

The perception on readability of 11 different coding style practices is tested by~\citet{dos2018impacts}. 
Using a custom tool, participants were shown a pair of code examples, one violating the coding practice while the other complied with it.
They find that 7 out of the 11 tested practices increase readability, 1 decreased readability, and the remaining 3 did not present statistically significant effects. 

\citet{ebert2017confusion} took a look at confusion in code reviews by examining the comments left by reviewers.
What they show is that reviewers often do not understand the context of the code change well, which has an impact on the understanding of the code.
However, reviewers are decently well in detecting and pointing out sources of confusion.




\section{Methodology}\label{cha:methodology}

The goal of this study is to measure the impact of atoms of confusion in Java code among computer science students.
To that aim, we propose the following research questions:
\begin{description}
    \item[RQ1] \RQone{}
    \item[RQ2] \RQtwo{}
\end{description}

\begin{figure}
    \centering
    \includegraphics[width=1\columnwidth]{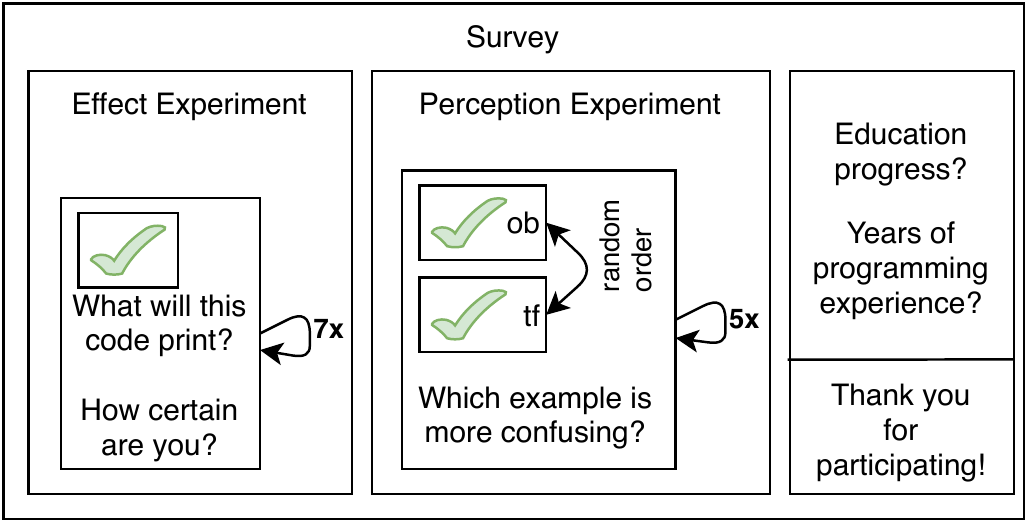}
    \caption{The design of the study. Participants first take the effect experiment (part 1, repeated 7 times) where they are presented with snippets that either contain or does not contain the atom of confusion and are asked to write down the output of the problem. In the perception experiment (part 2, repeated 5 times), participants are randomly presented with both versions of the snippet and are asked to evaluate which one they perceive as more confusing.}
    \label{fig:studyDesign}
\end{figure}

In Figure~\ref{fig:studyDesign}, we illustrate our methodology.
We first take the tasks from \citet{gopsteinSet2017}, that were originally devised for the C language, and manually translate them to Java code.
This process is described in Section~\ref{sec:atomsofconfusion}.
With the code snippets in hands, we devised a two-phase experiment.
The first part focuses on the effects of atoms of confusion. We give random code snippets to participants (some of them containing the atom of confusion) and ask participants to reason about the output of the program, which we then compare to the oracle answers.
In the second part, we show random code snippets containing the programs with and without the atom of confusion, and we ask participants which one they perceive as less confusing. We better describe the experiment in the following sub-sections.

\subsection{The Set of Atoms of Confusion}
\label{sec:atomsofconfusion}

We devise a set of code snippets containing different atoms of confusion in Java, based on \citet{gopstein2017understanding}'s work on atoms of confusion in C.\footnote{\url{https://atomsofconfusion.com/}}
The code snippets are, in essence, short programs with simple logic, and are affected by their respective atom of confusion.

We first explored which atoms have a Java equivalent.
This was not case for seven of the proposed atoms of confusion.
We did not translate the \textit{Implicit Predicate} atom, since Java's type system requires boolean values for predicates.
The type system also prevents the \textit{Pointer Arithmetic} and \textit{Assignment as Value} atoms from having a suitable Java translation.
In the case of \textit{Pointer Arithmetic}, a number cannot be added to a String type, and in the \textit{Assignment as Value} case, the inner assignment does not have a valid return type for the outer assignment.
The \textit{Macro Operator Precedence} and \textit{Preprocessor in Statement} atoms are not translated since preprocessing and macros are not part of the language specification.
Finally, the \textit{Comma Operator} and \textit{Reversed Subscripts} atoms cannot be translated as the syntax is not present, and we could not find similar behaviour in the language.

With the list of possible atoms of confusion in Java, we then translated the three variants (each containing the program with and without the atom of confusion) that \citet{gopstein2017understanding} devised for each of the atoms.
In order to stay in line with the original study, we make the programming language-related translations as similar as possible to the behaviour of the original source.

We ended up with a list of 14 atoms of confusion, which we show in Table~\ref{tab:atomsofconfusionJava}, and 80 different code snippets (14 atoms of confusion $\times$ 2-3 variants per atom $\times$ with and without the confusion). For two of the atoms, we could not translate one code snippet. 
We removed Gopstein et al.~\cite{gopstein2017understanding}'s code example number
52, related to ``Change of Literal Encoding'', as we argue that it unnecessarily requires participants to have knowledge about the ASCII character tables. We also removed code example number 59, related to ``Type Conversion'' because we did not find a trivial and small translation to an unsigned cast operation.
Thus, we ended up with 80 instead of 84 code snippets, all shown in our online appendix~\cite{appendix}.

\subsection{Study Design}
\label{sec:surveydesign}

\begin{table*}
\centering
    \caption{Atoms of confusion in Java, based on the work of \citet{gopstein2017understanding}}
        \begin{tabular}{lll}
            \toprule
            \textbf{Atom Name}                & \textbf{Java Code Snippet with Atom of Confusion} & \textbf{Java Code Snippet Free of the Confusion} \\
            \midrule
            Infix Operator Precedence         & \begin{lstlisting}[]
2 - 4 / 2\end{lstlisting}
            & \begin{lstlisting}[]
2 - (4 / 2)\end{lstlisting}
            \\
            Post-Increment/Decrement          & \begin{lstlisting}[]
V1 = V2++; \end{lstlisting}
            & \begin{lstlisting}[]
V1 = V2; V2 += 1; \end{lstlisting}
            \\
            Pre-Increment/Decrement           & \begin{lstlisting}[]
V1 = ++V2; \end{lstlisting}
            & \begin{lstlisting}[]
V2 += 1; V1 = V2; \end{lstlisting}
            \\
            Constant Variables          & \begin{lstlisting}[]
V1 = V2; \end{lstlisting}
            & \begin{lstlisting}[]
V1 = 5; \end{lstlisting}
            \\
            Remove Indentation atom           & \begin{lstlisting}[]
while (V2 > 0)
    V2--;
    V1++ \end{lstlisting}
            & \begin{lstlisting}[]
while (V2 > 0)
    V2--;
V1++ \end{lstlisting}
            \\
            Conditional Operator              & \begin{lstlisting}[]
V2 = V1 == 3 ? 2 : 1; \end{lstlisting}
            & \begin{lstlisting}[]
if (V1 == 3) { V2 = 2; } 
else { V2 = 1; } \end{lstlisting}
            \\
            Arithmetic as Logic         & \begin{lstlisting}[]
(V1 - 3) * (V2 - 4) != 0 \end{lstlisting}
            & \begin{lstlisting}[]
V1 != 3 && V2 != 4 \end{lstlisting}
            \\
            Logic as Control Flow             & \begin{lstlisting}[]
V1 == ++V1> 0 || ++V2 > 0; \end{lstlisting}
            & \begin{lstlisting}[]
if ( !(V1 + 1 > 0) ) 
    { V2 += 1;}
V1 += 1 \end{lstlisting}
            \\ 
            Repurposed Variables\textsuperscript{a}              & \begin{lstlisting}[,tabsize=2]
for(int V1 = 0;...; V1++) {
  for(int V2 = 0;...; V1++) { \end{lstlisting}
            & \begin{lstlisting}[]
for (int V1 = 0;...; V1++) {
  for (int V2 = 0;...; V2++) { \end{lstlisting}
            \\
            Dead, Unreachable, Repeated & \begin{lstlisting}[]
V1 = 1; V1 = 2; \end{lstlisting}
            & \begin{lstlisting}[]
V1 = 2; \end{lstlisting}
            \\            
            Change of Literal Encoding        &
            \begin{lstlisting}[]
V1 = 013 \end{lstlisting}
            & \begin{lstlisting}[]
V1 = Integer.parseInt("13", 8) \end{lstlisting}
            \\
            Omitted Curly Braces              & \begin{lstlisting}[]
if (V1) F1(); F2(); \end{lstlisting}
            & \begin{lstlisting}[]
if (V1) { F1(); } F2(); \end{lstlisting}
            \\
            Type Conversion                   & \begin{lstlisting}[]
V1 = (int) 1.99f; \end{lstlisting}
            & \begin{lstlisting}[]
V1 = (int) Math.floor(1.99f); \end{lstlisting}
            \\             
            Indentation                       & \begin{lstlisting}[]
if (V1 > 0) { } 
 V2 = 4 \end{lstlisting}
            & \begin{lstlisting}[]
if (V1 > 0) { }
V2 = 4 \end{lstlisting}
            \\
            \bottomrule
\multicolumn{3}{p{15.5cm}}{
{\textsuperscript{a}The \textit{repurposed variables} is the only atom where programs with and without it differ in semantics. While one may not consider it an atom of confusion (but a bug), we kept this atom as to better compare to the related work.}}
        \end{tabular}

    \label{tab:atomsofconfusionJava}
\end{table*}

The study consists of two parts, which will answer the first and second research questions, respectively.
In the first part of the study, we show participants code snippets, and ask them to evaluate the code and predict its output based solely on their thinking process.
In the second part, we ask participants to evaluate the code snippet, with and without the atom of confusion, choose the one they believe to be the least confusing, and explain why.

The experiment is conducted online, using the SurveyGizmo\footnote{\url{https://app.surveygizmo.com}} platform.
The experiment starts with an introduction to the study and an approximation of time needed.
We briefly explain the structure of the experiment and its guidelines.
We ask participants to fill in the answer solely based on their own knowledge, without the help of others and tools.
Furthermore, the guide specifies several points of information, based on the one used and provided by \citet{gopstein2017understanding}, and can be summarized as follows:
\begin{itemize}
    \item There is no time pressure, but the participant is encouraged to not stick to one question for too long.
    \item No syntax errors are present in the given code.
          If the participant does think errors are present, he/she is asked to explain in the comments where this error would take place.
    \item Encouragement not to use a calculator, the computer or a search engine for finding answers during the experiment.
    \item It is not possible to go back to previous questions.
    \item Discouragement of taking a break while filling in the survey.
\end{itemize}

Participants then work on the two parts of the experiment (the effects and their perception on atoms of confusion), which are detailed in the following two sub-sections.
After participants finish the two tasks, we collect their demographic information.
Lastly, the participant can optionally leave an email address to enrol for the possibility of winning a gift card.
The design of the survey was approved by the Human Research Ethics Committee of our university.

The estimated total time required for the experiment is 15 minutes, and consists of 12 tasks, seven for the first part and five for the second part. The entire experiment, code snippets, and datasets can be found in the online appendix~\cite{appendix}.

\subsection{Part 1: The Effect of the Atoms of Confusion}

The first part of the experiment is based on the study of \citet{gopstein2017understanding}.
We set out to understand whether the atoms of confusion hinder code understanding.

We show seven different code snippets, one at a time, to each participant.
Each code snippets shows a random program and asks the participant to write down what this program will print when executed.
We randomly pick a program from our database of code snippets; the code snippet may or may not contain the atom of confusion. Participants do not have this information.

In addition to the question on what the code will print, we ask participants how certain they are of their answer.
This question can indicate when the participant answers a question correctly while also being confused by the code, or even worse, when the participant answers a question wrong, believing he understood the code.
We ask this question in the form of a Likert scale~\cite{Jamieson2004}; the participant is asked to which degree they agree/disagree with the statement ``I am certain of the correctness of my answer above.''

Before the participant is directed to the first real question, we show one example exercise.
A pre-defined code example is given, with example answers already filled in for the questions.
The goal of the examples is to show the participant what is expected and to give an example of how the open questions can be answered.

Similar to the study by \citet{gopstein2017understanding}, we incorporate strategies to cope with the possibility of a learning effect~\cite{Neely1991}.
The code snippets we show to participants, and whether they contain the atom of confusion or not, are randomly chosen, distributing the bias of question order over the participants.
No specific actions are taken to prevent a single participant from seeing multiple variants of the same atom category, but it is guaranteed that pairs are only used for one question (e.g., the participant will not see the same snippet, with and without the atom).

\subsection{Part 2: Perception of the atom of confusion}
\label{ssec:part2}

In this part, we complement \citet{gopstein2017understanding}'s study design. We show participants two code snippets, side-by-side. Both snippets are the same program, but one of them contains the atom, and the other does not. Participants do not know what atom of confusion is presented to that code snippet, as well as which variant contains the atom of confusion and which does not. We show five pairs of snippets, one at a time, to participants.

We then ask which of the two code snippets the participant perceives as more confusing.
The four options are:
\begin{itemize}
    \item 1 is more confusing
    \item 2 is more confusing
    \item Both are equally confusing
    \item Neither are confusing
\end{itemize}
These four options provide the participant to add some granularity to their answer while still being easily categorizable.
While the third and fourth options do not distinguish the two variants, a difference in meaning is present in the options.
Answering ``Both are equally confusing'' would mean that both variants of this atom are not very readable code and can confuse developers, while answering ``Neither are confusing'' hints that both variants are not confusing or sufficient readable.

Participants are also asked to optionally explain their answer, enabling us to more precisely identify what the reason of the confusion is.
The confusion might be unrelated to the purpose of the atom, and would indicate that (the representation of) this atom is not suitable.

The code snippets we show to participants are also randomly selected from our dataset. We also ensure that the (also randomly selected) code snippets that a participant sees in Part 1 are not included in the set of code snippets that are randomly selected for Part 2.

\subsection{Data analysis}
\label{sec:dataanalysis}

For RQ1, the answers participants gave on the question ``What do you think this code will print?'' are compared to the correct answer for that code example.
We use the results to compute the odds ratio and confidence interval of wrong answers being caused by the atoms of confusion.
The corresponding confidence interval determines whether the confusion caused by the atom is statistically significant.
The odds ratio, its standard error and 95\% confidence interval are calculated according to \citet{AltmanDG2004}.
Given that zeroes may cause problems with computation of the odds ratio or its standard error, we add an extra 0.5 to all input values~\cite{Deeks2010,Pagano2000}.

We use the question regarding the certainty of the answer the participants to find out whether participants feel less certain in more confusing situations.
Additionally, the situation when the participant indicates high certainty of their answer while they answered incorrectly can be used as an indicator that this atom is important to be avoided in real life.

For RQ2, we use the perceptions part of the survey as input.
The participants' answers are collected and grouped per atom.
The results indicate to what extend the obfuscated atoms are perceived as being more confusing than their transformed counterpart.
Additionally, we qualitatively analyzed the participants' answers in the open question that asked them to reason about why they picked one option or another. We present these reasons, per atom, in the Discussion section of this paper.

\subsection{Participants}
\label{sec:participants}

We targeted the Computer Science students from our university as participants. We shared the experiment in the shared labs and in the internal communication tool.

At the end, 132 students took our survey. 
When participants do not finish the survey, the answers they filled in so far are still stored.
These responses will be marked as ``incomplete'', but the given answers might still be useful.
The incomplete responses with some answers filled in are extracted and used in the described analysis.

96 students took the complete study, and 36 students partially completed it. 
Out of the 96 participants that filled their demographic information, 58 (60\%) are 1st-year students, 14 (15\%) are 2nd-year students, 5 (5\%) are 3rd-year students, 14 (15\%) are MSc students. In addition, we have two teaching assistants and two students in the bridging program to the Computer Science Master's programme. One participant did not provide a valid answer.


\section{Results}
\label{cha:results}

\subsection{RQ1: \RQone{}}

\begin{table*}
    \centering
    \caption{The observed number of correct and incorrect answers, per atom of confusion, and their respective odds ratio with 95\% confidence intervals.
    The star (*) symbol indicates a statistically significant difference.}
    \begin{tabularx}{0.8\linewidth}{Xrrrrrrr}
        \toprule       
        
        & \multicolumn{2}{c}{\textbf{With the atom}} & \multicolumn{2}{c}{\textbf{Without the atom}} &  & \multicolumn{2}{c}{\textbf{Confidence}} \\
        
               & \multicolumn{2}{c}{\textbf{of confusion}} & \multicolumn{2}{c}{\textbf{of confusion}} &  & \multicolumn{2}{c}{\textbf{interval}} \\
        
        \cmidrule{2-3}\cmidrule{4-5}\cmidrule{7-8}
        \textbf{Atom}              & \textbf{Correct}                        & \textbf{Wrong}                           & \textbf{Correct}     & \textbf{Wrong}                                   & \textbf{Odds ratio} & \textbf{From} & \textbf{To} \\

        \midrule
        infix operator precedence  & 14  & 4   & 31  & 1  & 8.86            & 0.91 & 86.63  \\ 
        post increment decrement   & 14  & 17  & 32  & 3  & \textbf{12.95*} & 3.26 & 51.42  \\ 
        pre increment decrement    & 17  & 17  & 27  & 10 & \textbf{2.70*}  & 1.00 & 7.26   \\
        constant variables         & 30  & 0   & 27  & 0  & 0.90            & 0.02 & 47.00  \\
        remove indentation atom    & 16  & 17  & 26  & 0  & \textbf{56.21*} & 3.16 & 998.97 \\ 
        conditional operator       & 23  & 6   & 27  & 0  & 15.21           & 0.81 & 284.53 \\
        arithmetic as logic        & 24  & 4   & 41  & 0  & 15.24           & 0.79 & 295.42 \\
        logic as control flow      & 5   & 22  & 20  & 8  & \textbf{11.00*} & 3.09 & 39.21  \\ 
        repurposed variables       & 13  & 12  & 14  & 18 & 0.72            & 0.25 & 2.05   \\
        dead unreachable repeated  & 25  & 2   & 29  & 0  & 5.78            & 0.27 & 126.15 \\
        change of literal encoding & 4   & 16  & 12  & 10 & \textbf{4.80*}  & 1.21 & 19.08  \\ 
        omitted curly braces       & 19  & 13  & 27  & 4  & \textbf{4.62*}  & 1.30 & 16.36  \\ 
        type conversion            & 10  & 13  & 18  & 1  & \textbf{23.40*} & 2.66 & 206.16 \\ 
        indentation                & 31  & 2   & 24  & 0  & 3.89            & 0.18 & 84.78  \\
        \midrule%
        \textbf{Totals:}           & 245 & 145 & 355 & 55 & \textbf{3.82*}  & 2.69 & 5.42  \\\bottomrule
    \end{tabularx}
    \label{tab:effectResults}
\end{table*}

In Table~\ref{tab:effectResults}, we show the observed number of correct and wrong answers.

In total, 315 questions with an atom of confusion in the question code were answered, of which 115 received a wrong answer, opposed to 46 wrong answers on 337 questions with the atom translated out. 

These numbers indicate a statistically significant difference for seven out of the 14 tested atoms of confusion (i.e., \emph{post increment decrement}, \emph{pre increment decrement}, \emph{remove indentation atom}, \emph{logic as control flow}, \emph{change of literal encoding}, \emph{omitted curly braces}, and \emph{type conversion}). In other words, participants made significantly more mistakes in the code snippets affected by these atoms than in the code snippets not affected by them.

We also observe a statistically significant difference when we group all the atoms. We see an odds ratio of $3.82$ within a 95\% confidence interval of $[2.69, 5.42]$, indicating that, in general, code with atoms of confusion causes more understanding errors.


\subsection{RQ2: \RQtwo{}}

\begin{table}
\centering
    \caption{The number of participants that perceived the atom, the clean, both, or neither versions of code snippets confusing.}
    \begin{tabular}{lrrrr}
        \toprule

                      & \textbf{Atom}                                     & \textbf{Clean}                                         & \textbf{Both} & \textbf{Neither} \\
        
        \midrule
        infix operator precedence  & 18 (45.0\%)                                             & 5                                                   & 4             & 13               \\
        post increment decrement   & 23 (65.7\%)                                              & 2                                                   & 4             & 6                \\
        pre increment decrement    & 14 (41.2\%)                                             & 5                                                   & 7             & 8                \\
        constant variables         & 10 (27.8\%)                                             & 1                                                   & 0             & 25               \\
        remove indentation atom    & 20 (57.1\%)                                              & 5                                                   & 8             & 2                \\
        conditional operator       & 20 (60.6\%)                                             & 5                                                   & 0             & 8                \\
        arithmetic as logic        & 30 (71.4\%)                                             & 2                                                   & 3             & 7                \\
        logic as control flow      & 20 (55.6\%)                                              & 10                                                  & 2             & 4                \\
        repurposed variables       & 20 (47.6\%)                                             & 6                                                   & 13            & 3                \\
        dead unreachable repeated  & 23 (53.5\%)                                             & 0                                                   & 3             & 17               \\
        change of literal encoding & 5 (27.8\%)                                              & 8                                                   & 3             & 2                \\
        omitted curly braces       & 28 (93.3\%)                                             & 0                                                   & 1             & 1                    \\
        type conversion            & 8 (28.6\%)                                              & 5                                                   & 4             & 11               \\
        indentation                & 24 (72.7\%)                                             & 1                                                   & 2             & 6                \\\midrule
        \textbf{Total}             & 263 (54.2\%)                                             & 55                                                  & 54            & 113
                                   \\\bottomrule
    \end{tabular}
    \label{tab:perceptResultsPerQuestion}
\end{table}

We show the participants' perception regarding the atoms in Table~\ref{tab:perceptResultsPerQuestion}.

For 8 out of the 14 atoms (i.e., \emph{post increment decrement}, \emph{remove indentation atom}, \emph{conditional operator}, \emph{arithmetic as logic}, \emph{logic as control flow}, \emph{dead unreachable repeated}, \emph{omitted curly braces}, and \emph{indentation}), more than 50\% of participants agreed that the obfuscated variant is more confusing.
In other two atoms (i.e., \emph{infix operator precedence}, \emph{pre increment decrement}), while we observe most people still choosing the version with the atom of confusion to be more confusing, we observe some disagreement.

We also observe some atoms being considered less confusing. For the \emph{constant variables}, \emph{repurposed variables}, and \emph{type conversion} atoms of confusion, we observe that combination of neither or both being more popular among participants. Interestingly, the clean version of the \emph{change of literal encoding} atom was considered more confusing than the version with the atom of confusion itself.

\section{Discussion}

In Table~\ref{tab:summary}, we summarise our observations, per atom. 
We now discuss the consequences of our results, atom-by-atom.

\subsection{Infix Operator precedence}

The \textit{infix operator precedence} atom is all about the order of operations in single-line statements.
The results in Table~\ref{tab:effectResults} show that this atom is not significantly confusing.
The targeted confusion is caused by assuming an incorrect order of execution when more than one operator is used in the same line of code.
The transformed variant of the atom includes parenthesis around the operators to make the order of operations more straightforward.
In RQ2, 45\% of participants indicate the obfuscated variant to be more confusing.
The explanations from this group describe that the additional parenthesis improves readability by making the order of operations more clear.
One of the participants in the work of \citet{medeiros2018investigation} (in the research on atoms of confusion in open-source projects) states: ``[s/he] prefers to have parenthesis always, to [him/her] it makes it simpler to read''.
Another 42.5\% (combining the ``neither'' and ``both'' answers) states there is no difference between the two.
Arguments here state that the order of operations is clear, but knowledge of the precedence rules is needed.

12.5\% of answers given indicate the code without the confusing pattern to be more confusing.
All these answers came from one specific used code example where the precedence of the \texttt{!} (negation) operator was made more explicit by adding parenthesis.
The translation changes the order the operations are listed and adds optional parenthesis to indicate the order of operations.
In this example, as opposed to the other two examples, the parentheses are considered unnecessary and `making it look cluttered' or harder to follow/read by the participants. The same phenomenon is also observed by \citet{medeiros2018investigation}.

\newcommand{\faThumbsOUp}[0]{Yes}
\newcommand{\faThumbsODown}[0]{No}
\newcommand{\faLittle}[0]{little used}
\newcommand{\faCommonly}[0]{commonly used}

\begin{table}
\centering
    \caption{The summary of the results, per atom of confusion, in comparison to the results of \citet{gopstein2017understanding}}
    \resizebox{\columnwidth}{!}{
    \begin{tabular}{lccc}
        \toprule
& \multicolumn{2}{c}{\textbf{This work}} & \textbf{Gopstein et al.} \\
\midrule
\textbf{Atom of confusion} & \textbf{Confusion} & \textbf{Perception} & \textbf{Confusion} \\

        \midrule
        infix operator precedence  & \faThumbsODown{}   & \faThumbsOUp{} &  \faThumbsOUp{}    \\
        post increment decrement   & \faThumbsOUp{} & \faThumbsOUp{}  & \faThumbsOUp{}         \\
        pre increment decrement    & \faThumbsOUp{} & \faThumbsODown{} & \faThumbsOUp{}\\
        constant variables         & \faThumbsODown{}   & \faThumbsODown{} & \faThumbsODown{} \\
        remove indentation atom    & \faThumbsOUp{} & \faThumbsOUp{} & \faThumbsOUp{}          \\
        conditional operator       & \faThumbsODown{}   & \faThumbsOUp{} & \faThumbsOUp{}        \\
        arithmetic as logic        & \faThumbsODown{}   & \faThumbsOUp{} & \faThumbsODown{} \\
        logic as control flow      & \faThumbsOUp{} & \faThumbsOUp{} & \faThumbsOUp{}           \\
        repurposed variables       & \faThumbsODown{}   & \faThumbsOUp{} & \faThumbsOUp{}       \\
        dead unreachable repeated  & \faThumbsODown{}   & \faThumbsOUp{} & \faThumbsODown{}        \\
        change of literal encoding & \faThumbsOUp{} & \faThumbsODown{} & \faThumbsOUp{}            \\
        omitted curly braces       & \faThumbsOUp{} & \faThumbsOUp{}    & \faThumbsOUp{}          \\
        type conversion            & \faThumbsOUp{} & \faThumbsODown{} & \faThumbsOUp{} \\
        indentation                & \faThumbsODown{}   & \faThumbsOUp{} & \faThumbsOUp{}          \\\bottomrule
    \end{tabular}
    }
    \label{tab:summary}
    \vspace{-2mm}
\end{table}

\subsection{Post Increment Decrement}

The \textit{post increment operator} increments the variable and returns the original value of this variable.
Results in RQ1 show that this atom is significantly confusing.

The confusion can be caused by different misconceptions.
First, the operator might not be recognized.
However, only two of the participants that had any question regarding the post-increment atom indicated that they are not sure if the value of the variable will be changed or if the value of the original variable will be returned.
This shows that this operator is familiar to the participants.

A second possible misunderstanding is confusing the postfix increment/decrement operator with the prefix increment/decrement operator.
Instead of returning the original value, the pre-increment or decrement operator returns the result of the expression.
In the obfuscated variant of questions 8 and 9 (see appendix ~\cite{appendix}), the majority of explanations provided for wrong answers indicate that this is the cause of the answer being incorrect.
Another reason for the confusion is forgetting that the operator changes the variable.
One given answer indicates: ``\textit{\ldots{} Not sure if it will change the value \ldots}''.

Given these causes of confusion, and a significant result in Table~\ref{tab:effectResults}, with a lower bound of the confidence interval of 3.26, we argue that this atom is significantly confusing.
Moreover, 65.7\% of participants answer that the code examples with the atom of confusion are perceived as more confusing.
In the explanations, most participants indicate they know they are unsure about this atom.
A participant indicates ``\textit{I always forget where V1++ evaluates to (old or new value).}'' clearly describing what makes this participant confused.
Another participant answers ``\textit{Pretty sure that having the "++" after the variable makes sure it is evaluated before it is incremented, but I am still not sure and would probably have to just run the code to find out}'' making it even more clear that he/she knows where the confusion is and knows how to find the behaviour in the current situation.

\subsection{Pre Increment Decrement}

The \textit{pre increment decrement} atom is very similar to the post-increment decrement atom.
The difference, as explained before, is that instead of the original value of the variable, the result of the expression is returned.

In contrast to the previous atom, participants indicate they do not know what \texttt{{-}{-}V1} would do.
Five participants state they are unsure about or are unfamiliar with the syntax.
The transformed variant of this atom does not remove the operator, but instead isolates it.
This removes the confusion with the behaviour of prefix operator, but does not resolve confusion caused by syntax.

The first experiment shows that this atom is confusing.
The confidence interval starts ever so slightly above 1. 
The incorrect answers include ``\textit{Idk if ++V1 is a thing. I always use V1++}'' and ``\textit{I have no clue what "{-}{-}v1" means}''.
This shows that these people are confused by the atom.
The second experiment confirms this finding.
The participant answering ``\textit{I can never remember what {-}{-}V1 really does so to me this is confusing}'' clearly indicates that this atom is causing confusion.
41.2\% of the participants agree that the code variants with the atom of confusion included are more difficult to understand.

One notable observation by participants is that the transformed variant in the code snippet 10 (in our appendix) initializes two variables in the same line: \texttt{int V1 = 5, V2;}.
This causes unnecessary confusion since some participants indicated this as the reason for confusion. We suggest future replications to improve the code snippets for this atom of confusion.

\subsection{Constant Variables}

The \textit{constant variable} atom was not shown to be statistically significant in the research by \citet{gopstein2017understanding}.
This research confirms this result.
None of the participants made a mistake with this atom.
The only slight confusion indicated by the participants was the uncertainty whether the printed text would include the \texttt{.0}, caused by the \textit{double} type of the variable.
We considered answers not including the \texttt{.0} to be correct since this is not part of the confusion under study.
69.4\% of the answers in the second part of the survey indicate no difference in confusion between the variants of the questions for this atom.
The majority of the explanations state that the code is very simple to be understood.
The main reason why the obfuscated atom would be considered more confusing is caused by the unnecessary extra code.
The added verbosity contributes to complexity in the used code examples.
Future research should investigate the effects of this pattern when present in larger code snippets.

\subsection{(Remove) Indentation Atom}

Along with the atom \textit{indentation}, these two atoms are included in this research despite being removed in the original experiment by \citet{gopstein2017understanding}.
In their errata\footnote{\url{https://atomsofconfusion.com/2016snippetstudy/
errata.html}}, authors state that: ``\textit{To remove the bias introduced by code formatting, we chose not to study the effect of whitespace in this study}''.
This study did include these two atoms to explore the impact of misformatted code.
The results show that the remove indentation atom is indeed significant confusing.
The lack of indentation makes it difficult to see where scopes are ending, resulting in a majority of wrong answers for the obfuscated variants.
All participants that encountered a variant without the atom in the first part of the survey gave the correct answer.

Interestingly, 
in the \textit{indentation} atom, the brackets are included, and only the indentation itself is wrong or missing. That atom is not significant confusing, with only two wrong answers for the obfuscated variants.

One of our code snippet variants is an example of the ``dangling else'' pattern as researched by \citet{medeiros2018investigation}.
Authors state that many coding standards enforce the use of brackets to avoid this pattern.
Our results cohere with this conclusion, showing major confusion with missing indentation when brackets are left out and no confusion when the brackets are included.

\subsection{Conditional (Ternary) Operator}

In Java, the available ternary operator is the \texttt{?\ :} operator, which provides a shorthand way of writing an \texttt{if-then-else} statement.\footnote{\url{https://docs.oracle.com/javase/tutorial/java/nutsandbolts/op2.html}}
Although 20\% of the answers for the variants containing this atom were wrong, and only correct answers were given to the transformed variants, we do not have enough data to draw significant conclusions.
The participants that provided incorrect and/or were less sure on their answers state that they are unsure about the syntax.
The notes left by participants that correctly answered the code snippet with the atom show they are unsure about the functioning of the operator, but have a correct intuition.
Explanations regarding the second research question indicate that a major part of confusion for the obfuscated variants comes from the lack of parenthesis around the condition, e.g., \texttt{V2 = (V1 == 3) ? 2 : 1;}.

A large number of participants indicate that the ternary operator makes code unclear.
Others state that it can be confusing to people unfamiliar with the operator.
On the other hand, participants that preferred the if ternary instead of traditional if-else statements argue that the if ternary reduces the verbosity of the code; according to them, having to read or write 4 or more lines of code instead of 1 makes the understanding take longer when familiar with both syntax variants.
\citet{kernighan1999practice} state in their book that using the ternary operator to replace four lines of if-else code is a good idea. Interestingly, \citet{medeiros2018investigation} decided not to investigate the effects of this atom given its popularity in real-world codebases.

\subsection{Arithmetic as Logic}

Similarly to \citet{gopstein2017understanding}, our research shows that this atom does not confuse developers.
The main assumption from \citet{gopstein2017understanding} is that using arithmetic operators instead of logical operations, will imply a non-boolean range, which might be confusing.
Due to the translation to Java, the resulting number has to be explicitly compared to 0 to create a boolean value, taking away this implication of a non-boolean range. 
The results in the second part of our experiment show that the variant with the atom of confusion is much less preferred to read.
The additional calculations that are unusual lead to a much longer time to see the intention of the code, according to the participants.
Moreover, we observed some complaints about the order of comparison in the translated variants of this atom.
Reversing the order of \texttt{variable <comparison operator> value} is disliked by one of the participants.
The effect of the order reversal is interesting for future research.

\subsection{Logic as Control Flow}

Due to lazy evaluation of the \(||\) and \texttt{\&{}\&{}} logical operators, they can also be used as conditional operators.
This means that, depending on the value of the left side, the right sight may or may not be executed.
The most significant single result for this atom in part 1 of the experiment is that one of the variant with the atom (code snippet 41 in our appendix) was answered incorrectly by all the participants.
In total, the variants with this atom have an error rate of 81\%; participants were 11 times more likely to make mistakes in code containing this atom, in contrast to code not containing it.

When comparing code snippets with and without the atom, participants affirmed that:
``\textit{Using logic expressions to perform effect-full computation is cool [..], but more confusing because it breaks the vertical flow of effect-full expressions/statements.}'' and ``\textit{++variables should be replaced with something more clear}''. Interestingly, we also observed participants defending the atom: ``\textit{I really dislike nesting whiles and ifs as done in [transformed variant]. You cannot easily trace it back yourself. [Obfuscated variant] is not perfectly readable, but it only has one check which is easier to do by hand.}''

\subsection{Repurposed Variables}

The \textit{repurposed variables} atom ``misuses'' an already existing variable for another purpose.
An exceptional result for this atom is that the variants where code examples did not include the atom of confusion are answered incorrectly more often than the variants with the atom of confusion.
For the variants with the atom of confusion, 13 participants answered correctly and 12 gave an incorrect answer.
The variants without the atom of confusion resulted in 14 correct and 18 incorrect answers.

The results of the perception part of the experiment fall more in line with the expected results.
47.6\% of participants agreed that the variant with the atom of confusion is indeed the more confusing.
A substantial amount of participants indicated that they found both variants to be confusing (38.1\%).
This is more in line with the findings in the first part of the experiment.

\subsection{Dead, Unreachable, Repeated}

This is one of the atoms that did not cause significant confusion, according to \citet{gopstein2017understanding}. In the case of our experiment, only two wrong answers were given, out of the 56 times this atom was present in the first part of the experiment, which seems to be inline with previous findings.

The shared factor in the three terms of this atom's name is that the removal of the corresponding line(s) of code will not change the behaviour of the program. 


\subsection{Change of Literal Encoding}

This atom focuses on the general case when the encoding of the characters written down will change in the meaning of the program, For example, prepending a number with a 0 tells Java to parse this number in the octal numbering system, e.g., 013 results in the decimal value of 11.

We observe a statistically significant difference between code snippets with and without atoms. Interestingly, in one of the code snippets, we observed participants answering this question quite fast, in terms of time.
This indicates that participants may quickly jump to a wrong conclusion in code containing this atom of confusion.
On the other hand, we observed that, when faced with the code snippet with and without the variants, participants chose the clean version of the snippet as more confused. This might indicate that, while this atom may cause confusion, developers should carefully decide how to write the alternative implementation.

\subsection{Omitted Curly Braces}

This atom is similar to the \textit{Remove Indentation Atom} and the \textit{Indentation} atom in the sense that the targeted confusion is related to unclear separation between code blocks.
Code snippets affected by this atom do not contain the curly braces (i.e., \texttt{\{ ... \}}) that normally follow if statements, while loops, and other branch instructions, which clearly mark the beginning and the end of that code block.
One participant identifies the confusion as: \textit{``Putting several statements on one line is unnecessarily confusing.''}
Another participant says: \textit{``I thought Java only executes the next statement after a for-loop without brackets but it also could be that the whole next line is executed.''}

41\% of participants that were asked to evaluate a task with this atom of confusion in the code answered incorrectly. This atom is significant confusing ($4.62$ times more likely to make mistakes in code with the atom). 
Moreover, a stunning 93\% of participants agree that it is more confusing to omit the curly braces from these code blocks.

\subsection{Type Conversion}

The \textit{Type Conversion} atom is about converting one type into another, e.g., \texttt{(int) 1.99f;}.
This atom was the one where participants had the most uncertainty when answering.
The particular code snippet regarding casting to the byte primitive type is especially often answered wrong.

Our results show that this atom causes significantly more confusion than the version without the atom. When presented with both snippets of code,
28.6\% of the participants indicate that the code examples including the atom of confusion are indeed harder to understand.
The argument given is that the code free of the atom is much more explicit in what is happening:
``\textit{You would have to know what happens if a float is cast to an int}''.
Interestingly, a large portion of participants also indicates that neither of the code variants is confusing.
A participant states that: ``\textit{I believe that casting to an integer automatically floors the value, while it might be easier to understand by flooring as well, I don't think adding redundant code makes it less confusing. While the second simply casts to an integer, flooring it as well.}''

\subsection{Indentation}\label{ssec:indentation}

Only two (out of 33) participants answered incorrectly.
As for the perception of the participants towards this atom, the majority agrees that the variants with the atom present are more confusing.
Furthermore, the explanations heavily gear to the intended explanation of this atom.
The comments often mention indentation or formatting, and how this affects the clearness of the distinction between different branches.
One participant provides an interesting reason for this that ``\textit{if you're skimming the code you could misread and think the [line of code] was in the if statement}''.

These results indicate that this atom of confusion is easily recognized, and therefore will not often cause actual misunderstanding.
Participants, however, have a strong preference to avoid this atom.

\section{Recommendations and Future Work}



Our results indicate that atoms of confusion may hinder the developers' abilities to comprehend the code; in particular, novice developers. Given that software development teams are often composed of developers with mixed levels of experience, we recommend teams to either avoid writing code that contains any of the atoms of confusion we observed to be significantly confusing, or to make sure developers of all levels are trained to understand them.

Educators can play an important role in making sure that the future generations of software developers suffer less from these atoms of confusion. We suggest educators to incorporate atoms of confusion in their teaching materials. The tasks that are used in this experiment can serve as examples. Interestingly, computer science education researchers have long been studying programming misconceptions and mistakes that novice developers make (e.g.,~\cite{caceffo2019identifying,brown2017novice,qian2020teachers,kaczmarczyk2010identifying}). However, the intersection between the misconceptions observed in these studies and the atoms of confusion that are currently studied by the software engineering community appears to be small. In fact, the atoms of confusion we study here do not appear in any of the educational papers we cite. We suggest both communities to work together in finding ways to improve the way developers learn about such code constructs.


Finally, there is still research to be done in the code comprehension of atoms of confusion. We suggest researchers to investigate: 
(i) the impact of these atoms of confusion among more experienced developers, i.e., do they also affect experienced developers, or do they only affect more novice developers?
(ii) the impact of atoms of confusion in larger code snippets; while larger code snippets will reduce the control of the experiment, it would better reflect the real-world scenario where those atoms are embedded in larger methods,
(iii) other atoms/code constructs that may cause confusion; the list of atoms from \citet{gopstein2017understanding} originated from an obfuscation competition; diving into other possible atoms of confusion, that were not necessarily created with the purpose of confusing,
(iv) given that the atoms we study are based on the atoms from Gopstein et al, we may have not studied atoms of confusion that are specific to the Java language, which can be the focus of future studies, and
(v) while most of our results matched with the ones from Gopstein et al~\cite{gopstein2017understanding}, some did not; an interesting line of inquiry would be to understand why some atoms are perceived as confusing in one language but not in another.

\section{Threats to Validity}


\textit{Threats to Construct and Internal Validity.} 
The tasks (i.e., code snippets) were inspired by the question set used by \citet{gopstein2017understanding}. 
In their work, the subjectiveness of making this set is mentioned as a threat to validity, which this work naturally inherits. 

Moreover, all code examples used in the experiment use short, non-descriptive identifier names like \texttt{V1, V2}.
Non-descriptive identifier names reduce comprehension of source code~\cite{schankin2018descriptive}.
Since the code examples are isolated, it is hard to describe the meaning of the individual variable names, and finding names that will not introduce additional bias between the different code examples will be even harder.
However, this does not rule out the effect the identifier names have on the comprehensibility of the source code atoms.

Finally, we note that the same participant provides data points for different atoms of confusion. We opted for not performing any type of correction (e.g., Bonferroni) for multiple comparisons. As suggested by Armstrong~\cite{armstrong2014use}, given that we are not testing an universal hypothesis (i.e., we test each atom of confusion separately) and that our study does not require all tests to be significant in order to validate the main hypothesis, correction is not required. Nevertheless, all the data is available in our appendix~\cite{appendix} for researchers to experiment with different forms of correction.


\textit{Threats to External Validity.} 
This study makes use of computer science students as participants. 
Our Computer Science and Engineering curriculum starts with a course on object-oriented programming (OOP), where the used language is Java.
After that, the students took part in an OOP project course, requiring the students to build an application in Java.
While it is unsure if all participants passed the course, they have been in contact with the Java programming language for a minimum of half a year.
This gives us a strong indication of the minimal amount of experience for every participant. Nevertheless, we do not argue that our findings are generalisable to any novice developers, computer science students, or the population of developers in general. Replications of this work are required.

Finally, as we mentioned before, we evaluate a set of atom of confusion that was initially proposed by \citet{gopstein2017understanding}. We do not argue that this set is complete and covers all possible atoms of confusion for the Java language. Expanding the set of atoms of confusion is part of our future agenda.

\section{Conclusions}
\label{cha:conclusion}

Code that is easy to comprehend is fundamental for sustainable software maintenance. 
In this paper, we explored the effects of atoms of confusion by measuring the reactions and impact of these confusing patterns in Java code among novice developers.

Our results show that (i) participants are 4.6 up to 56 times more likely to make mistakes in code snippets affected by 7 out of the 14 studied atoms of confusion, and (2) when faced with both versions of the code snippets, participants perceived the version affected by the atom of confusion to be more confusing and/or less readable in 10 out of the 14 studied atoms of confusion.

In other words, our results show that atoms of confusion can cause confusion among novice software developers. We suggest software companies to investigate how often these atoms happen in their codebases, and how much they actually impact in the productivity of their developers.

Finally, given that this paper shows that some atoms can indeed be confusing to Java developers, we suggest future research to explore other language-specific atoms. In addition, the Java community has long been proposing different code idioms, guidelines, and best coding practices. Future work should explore the pros and cons of such guidelines from a program comprehension perspective.

\section*{Acknowledgments}

We thank all the 132 computer science students that took their time and participated in this research.

\renewcommand*{\bibfont}{\footnotesize}
\bibliographystyle{IEEEtranSN}
\bibliography{paper.bib}

\end{document}